**Polymer-stabilized graphene dispersions at high concentrations in organic solvents for nanocomposite production**

Ahmed S. Wajid[a], Sriya Das[a], Fahmida Irin[a], H.S. Tanvir Ahmed[b], John L. Shelburne[a], Dorsa Parviz[a], Robert J. Fullerton[a], Alan F. Jankowski[b], Ronald C. Hedden[a], Micah J. Green[a*1]

*[a]Department of Chemical Engineering, [b]Department of Mechanical Engineering, Texas Tech University, Lubbock, Texas 79409, USA*

**Abstract**

We demonstrate a simple and effective technique for dispersing pristine (unfunctionalized) graphene at high concentrations in a wide range of organic solvents by use of a stabilizing polymer (polyvinylpyrrolidone, PVP). These polymer-stabilized graphene dispersions are shown to be highly stable and readily redispersible even after freeze-drying. This technique yields significantly higher graphene concentrations as compared to prior studies. Excellent enhancement in thermal conductivity of the fluid by addition of pristine graphene is also demonstrated. These well-dispersed pristine graphene sheets were then used as a strong and conductive nano-filler for polymer nanocomposites. PVP/graphene nanocomposites were produced via bulk polymerization of N-vinylpyrrolidone loaded with dispersed graphene, resulting in excellent load transfer and improved mechanical and electrical properties.

**1. Introduction**

Graphene, a single layer of sp$^2$ bonded carbon atoms, has attracted significant scientific importance in recent years because of its remarkable properties[1]. It is electrically more conductive than any other substance known, its mechanical strength is 200 times greater than that of steel at a sixth of the weight, and it has a thermal conductivity greater than that of diamond [2-4]. Graphene was initially isolated through micromechanical cleavage of graphite also known as the 'Scotch tape' method; this discovery was recently honored with the 2010 Nobel Prize in Physics[5]. Recent advancements in the processing of graphene include epitaxial growth, chemical vapor deposition (CVD) [6], and liquid phase exfoliation and processing [7]. The latter method is best suited for scalable production of multifunctional advanced materials such as graphene based nanocomposites [8], thin conductive films [9], thermal pastes, and ultra capacitors[10], as graphene needs to be exfoliated from graphite and dispersed in a liquid prior to its processing.

---
1  corresponding author, micah.green@ttu.edu, (806)742-5158

However, the liquid-phase production of graphene-based materials is severely hampered by the poor solubility of graphene in common organic solvents, which is chiefly due to the attractive van der Waals forces between graphene sheets. Even after the initial process of exfoliation and dispersion, van der Waals forces cause graphene to reaggregate.

The most common technique used in liquid phase dispersion of graphene is to oxidize the graphite to form graphite oxide [11-13], which may then be exfoliated to create single sheets of graphene oxide (GO). The oxidized functional groups render GO soluble in water. GO can be reduced to chemically converted graphene (CCG), but reduction does not completely restore the pristine graphene structure. Restacking also occurs during chemical reduction, which significantly reduces the effectiveness of this method [14]. This reaggregation of CCG can be prevented by using surfactants or polymers that act as stabilizers [15]. Although these stabilizers sterically prevent reaggregation [12, 13], this approach has its own limitations. During the process of reduction of GO to CCG, some of the $sp^3$ characteristics of GO are retained in the CCG [11] such that the CCG obtained does not have the properties as that of pristine graphene. For instance, the electrical conductivity of CCG is two orders of magnitude lower than that of pristine graphene [16]. Furthermore, recent work has suggested that GO is not actually soluble in water and other solvents; instead, oxidative debris acts as a stabilizer [17]. For these reasons, the exfoliation and dispersion of pristine graphene is pursued in this study.

Studies of pristine graphene dispersion (without oxidation) have largely focused on aqueous solutions, where graphite may be sonicated in the presence of surfactants such as sodium chlorate or sodium dodecyl benzene sulfonate (SDBS) [18-27]. However, such surfactants are typically undesirable in nanocomposite applications [28], and many nanocomposite processing techniques require the use of organic solvents. Techniques to disperse pristine graphene in organic solvents through sonication have been explored [7, 29] but suffer from certain drawbacks such as comparatively low concentrations and the need for extensive sonication. Although superacids act as an excellent solvent for graphene at concentrations as high as 2 mg/ml, superacids are typically incompatible with most nanocomposite applications [30].

However, the use of polymer stabilizers rather than micelle-forming surfactants has shown promise in stabilizing dispersed graphene. In particular, Bourlinos *et al.* used polyvinylpyrrolidone (PVP) as a "coating" polymer to prevent graphene aggregation in aqueous dispersions. This approach uses only sonication and avoids oxidation or other covalent functionalization [19]. PVP is a linear homopolymer of N-vinylpyrrolidone (VP) that is soluble in water and a number of other polar solvents, and prior studies indicate that PVP has a propensity to coat graphite surfaces [31]. PVP is inexpensive, biocompatible, and has many applications in various fields such as pharmaceuticals, cosmetics, food, adhesive, polymers, and textiles [32, 33].

The dispersion of graphene in solvents holds importance for the processing of reinforced polymer nanocomposites. Such nanocomposites show tremendous potential for mechanical properties enhancement due to high specific surface area, strong nanofiller-matrix adhesion, and the outstanding mechanical properties of the $sp^2$ carbon bonding network in graphene. Difficulties in dispersion of pristine graphene have limited the incorporation of pristine graphene into bulk nanocomposites. Thus, most of the prior graphene nanocomposite work has utilized GO and CCG [15, 34-42]. A few studies have addressed processing of pristine graphene into nanocomposites [7, 43, 44]. Recently, Vadukumpully *et al.* reported thin nanocomposite films produced from surfactant-wrapped graphene dispersions in DMF mixed with solution-cast poly(vinyl chloride)[45].

In the present investigation, high-concentration PVP-stabilized pristine graphene dispersions are demonstrated in a variety of solvents of industrial relevance. These dispersions are then employed in the manufacture of PVP/graphene nanocomposites. This work represents the first case of pristine graphene dispersion in a polymerizable monomer to form a two-component nanocomposite in a single step.

## 2. Experimental methods

**Materials:**

Expanded graphite (EG) (grade-3805) was purchased from Asbury Carbons. Poly(vinyl pyrrolidone) (PVP) (M.W: 10,000 g mol$^{-1}$) was bought from Sigma Aldrich. Vinyl Pyrrolidone (VP), Ethanol, Methanol, Dimethylformamide (DMF), Dimethylsulfoxide (DMSO), and N-methyl-2-pyrrolidone (NMP) were purchased from Sigma Aldrich and were used as purchased.

**Methods:**

Stable pristine graphene dispersion:   A PVP stock solution was prepared by magnetically stirring 10 mg/mL of PVP in a solvent for 10 minutes. Stable graphene dispersions were prepared by adding 40 mg/mL of Expanded Graphite (EG) to the PVP solution. The EG was then mixed with PVP solution for the same amount of time. The solution was then tip sonicated in a water bath using a Misonix sonicator (XL 2000) at output wattage of 10W for 1 h at room temperature. The dispersions were then centrifuged at ~ 5000 rpm for 4 h to sediment large aggregates or remaining EG, and the supernatant liquid was retained.

Freeze drying:  Polymer-stabilized graphene dispersions in water were freeze dried using a Vitris Benchtop Freeze Dryer over a period of 24 hours to obtain gray colored powder. The freeze dried PVP-stabilized graphene was re-dispersed in various solvents without any need of tip sonication.

Vacuum filtration: A regular vacuum filtration setup was utilized to measure the concentration of graphene in the dispersions. A polytetrafluoroethylene filter paper with a pore size of 0.02 μm was used. The mass of the filter paper was measured before and after filtration and was used to determine the concentration of graphene in the dispersion.

$$c_{graphene} = \frac{\text{Initial filter paper mass} - \text{final filter paper mass}}{\text{volume of the solvent}} \qquad (1)$$

Absorbance: To measure the concentration of graphene ($c$) in the dispersions and correlate with absorbance ($A$), UV-Vis spectroscopy was performed on a Shimadzu UV-Vis spectrophotometer 2550 at wavelengths of 200 nm to 800 nm. The samples are controllably diluted such that the measured absorbance is never above 2 because the equipment absorbance limit is approximately 4. The extinction coefficient of PVP stabilized graphene in various organic solvents was determined using the Lambert-Beer law.

$$A = \alpha l c \qquad (2)$$

This law states that the extinction coefficient ($\alpha$) of any substance varies linearly with the product of the concentration (c) and the path length (l). The absorbance of the dispersion was measured at a wavelength of 660 nm; at this wavelength, dispersed PVP has negligible effects. All absorbance measurements were measured against a blank of the appropriate PVP/solvent mixture.

<u>Raman</u>: Raman spectra were measured by a Renishaw Raman microscope using a 633 nm He-Ne laser. The graphene-coated polytetrafluoroethylene filter paper specimen obtained from vacuum filtration of the graphene dispersion was used as the sample for Raman spectroscopy.

<u>Thermal conductivity measurement:</u> The thermal conductivity of PVP-stabilized graphene dispersions was measured using KD2 Pro Thermal Properties Analyzer at 22 °C. This device uses the transient hot wire thermal conductivity technique [46].

<u>PVP nanocomposite preparation:</u> The PVP-stabilized graphene dispersion in VP was used as the starting material for making graphene-based PVP nanocomposites. 0.1 wt % of a photoinitiator (Irgacure 651, Ciba) was added to the graphene/PVP/VP dispersion to promote the VP polymerization. The solution was poured into quartz vials (inner diameter of 3 mm and length of 100 mm) and was subjected to UV light (High intensity UV-lamp, 100 W, 365 nm UV) at a distance of ~10 cm in a nitrogen atmosphere. Note that control (graphene-free) samples were prepared in the same manner, with VP polymerized in these quartz vials.

<u>Scanning Electron Microscopy:</u> SEM samples were prepared by mounting the samples on double-faced carbon tape and sputter coating with gold at 10 mA current for 1 min. An accelerating voltage of 2 kV was used to image the specimens on a Hitachi S4300 SE/N**.**

<u>Transmission Electron Microscopy:</u> TEM samples were prepared by depositing liquid samples on 400 mesh carbon-coated copper grids (Electron Microscopy Sciences, CF400-Cu) and air drying for 1 min. A voltage of 100 kV was used to image the samples on Hitachi H8100.

<u>Tensile Tests</u>: The tensile test specimens were mounted on a Test Resources universal testing machine using detachable clamps with serrated grip surfaces. Monotonic tensile testing was done on the specimens by moving the linear actuator of the machine over the displacement of 20 mm over 2000 sec. The data acquisition system logged the normal load from a load sensor as the displacement sensor (Linearly Variable Differential Transducer, LVDT) recorded the crosshead position of the actuator as a function of time at a user-specified frequency. The displacement-measured load curves were fit with linear trend-line to determine the elastic modulus and the deviation from the elastic regime.

Electrical conductivity measurements: The electrical resistance measurements were made by standard Four-point probe method. The four-point probe head (Signatone, SP4-40045TBY) consists of four equally spaced tungsten tips. Each tip was supported by springs to minimize sample damage during probing. The four-point probe head was mounted on a resistivity measurement stand (Signatone, Model 302). A high impedance current source (Keithley 2400) was used to supply current through the outer two probes; a voltmeter (Keithley 2000) measures the voltage across the inner two probes to determine the sample resistivity. The spacing between the probes is s ~ 1 mm.

## 3. Results and discussion

### 3.1 Graphene dispersion

In order to create high-concentration dispersions of pristine graphene in organic solvents, we utilized PVP as a stabilizer as it has been previously shown to stabilize graphene dispersions in water [47]. PVP prevents graphene reaggregation sterically by adsorbing on the graphene surface. We tip sonicate expanded graphite in a PVP solution for approximately 1 hour to achieve exfoliation. Tip sonication is far more efficient at exfoliating graphene as compared to bath sonication, which may take many hours to achieve effective exfoliation. The resulting dispersion was then centrifuged to remove any aggregates, and we measured the concentration of the supernatant through both vacuum filtration and absorbance measurements.

PVP effectively stabilizes the graphene in a variety of solvents, and the resulting dispersions are stable against centrifugation (Figure 1). Graphene flakes without the PVP reaggregate and sediment out during centrifugation whereas the graphene dispersed with PVP forms a stable solution. Thus, by using PVP, we are able to produce stable graphene dispersions in various solvents without any chemical modification to the graphene. Effective solvents included DMF, NMP, ethanol, methanol, VP, and dimethyl sulfoxide (DMSO) (Table 1, Figure S1, S2). Additional solvents were investigated and are listed in Table S1. The solubility of PVP in a given solvent has a major impact on the dispersibility of PVP-stabilized graphene in that solvent. Naturally, solvents that do not dissolve PVP were ineffective. Also, several solvents that do dissolve PVP, such as hexafluoroisopropanol (HFIP), chloroform and isopropanol, were unable to form stable graphene dispersions. This observation could be attributed to solvent-graphene surface energies, which impact the ternary graphene/PVP/solvent system as well. The dependence of the supernatant concentration of graphene on the PVP concentration is shown in Figure 2. Beyond a PVP concentration of 10 mg/mL, the graphene concentration does not increase significantly; this is similar to findings for other dispersed stabilized nanomaterials [48].

Without PVP, none of these solvents, including DMF and NMP, were able to consistently disperse pristine graphene. This observation conflicts with prior reports that pristine graphene may be dispersed directly in organic solvents such as DMF and NMP without the use of a stabilizer [49]. The same graphite source was used in both studies. These conflicting findings may be explained by the role of centrifugation. The centrifugation scaling proposed by Khan *et al.* indicates that the dispersions are not stable against prolonged centrifugation [50]. In contrast, for PVP-stabilized graphene dispersions, we find that concentration does not significantly decrease after 2-3 h of centrifugation at 5000 rpm (Figure 3), which is comparable to findings for surfactant-stabilized carbon nanomaterials [51].

The stability of PVP-stabilized graphene dispersions is also supported by freeze-drying experiments. A PVP/graphene/water dispersion was freeze-dried to yield a dark gray powder, which was easily re-dispersible in the solvents listed in Table 1 without the need for sonication. No sedimentation of graphene was seen after centrifugation (Figure 4), indicating that the PVP remains adsorbed on the graphene surface even after lyophilization, sterically preventing the graphene sheets from van der Waals contact. Furthermore, the PVP-stabilized graphene/water dispersions were found to be stable at higher temperatures ($\sim$100$^o$ C) and at low pH ($\sim$2), unlike surfactant-stabilized dispersions.

The effect of molecular weight of PVP on the graphene dispersion was also investigated. PVP stabilizers with three different molecular weights (10,000, 40,000 and 110,000 g mol$^{-1}$) were compared. The initial concentration of PVP was kept constant at 10 mg/mL, and the changes in the final concentration of graphene were observed (Figure 5). As expected, the effectiveness of PVP decreases with increase in molecular weight of PVP. This downward trend is likely due to increases in entanglement points per chain and decreased chain mobility as observed in prior studies [52].

To confirm the presence of single-to-few layer stabilized graphene in our dispersions, we characterize our samples using HRTEM. A HRTEM image of PVP-stabilized graphene deposited from ethanol is shown in Figure 6. Folded few-layer graphene sheets are visible in Figure 6(a). The images of the sheet edges in Figure 6(b), (c), and (d) indicate that the sheets are 2-4 layers thick. Additional HRTEM images are shown in Figure S3. Also, Raman spectroscopy was utilized to characterize the degree of exfoliation of graphene in the solvent. As seen in Figure S4, the shift in G-peak of PVP stabilized graphene dispersion in DMF as compared to expanded graphite indicates the exfoliation of graphene [53].

The enhancement in thermal conductivity of the solvent by the addition of graphene was measured and the results are summarized in Table 2. As seen in Table 2, with a very low loading of pristine graphene, the thermal conductivity of the fluid is enhanced by a factor of 1.5. The enhancement ratio of pristine graphene is much higher than that of CCG [54] and GO [55] at a much lower loading.

**3.2 PVP nanocomposites**

As PVP-stabilized graphene may be dispersed in N-vinylpyrrolidone monomer (VP), we investigated the bulk ultraviolet (UV) free-radical photopolymerization of VP to form PVP/graphene nanocomposites. The newly created polymer matrix and the stabilizer polymer are the same, allowing for excellent load transfer between the graphene and the matrix. This particular graphene/polymer nanocomposite is a suitable model system for comparison against theory for mechanical reinforcement because of this excellent load transfer.

The graphene/PVP/VP dispersion is poured into thin quartz cylinders and polymerized (Figure 7). Because the reaction is UV-catalyzed, the dark graphene can inhibit polymerization in thick samples; to combat this problem, we focused on thin, cylindrical samples inside quartz capillaries.

Tensile test results for baseline PVP and 0.03 vol % graphene/PVP nanocomposites are shown in Figure 8 and compared against theory. PVP is glassy at this temperature; viscoelastic effects are negligible [56]. For the theoretical calculations, we utilized the well-known Halpin-Tsai equations for composites with randomly oriented fillers [57-59]. Details of the calculation are described in supplementary information. The theoretical calculations predicted that the Young's modulus of the 0.03 vol % loaded graphene PVP nanocomposite should be 619 MPa, i.e. an increase of 24% from the control PVP nanocomposite that has a Young's modulus of 500 MPa. The tensile test results indicate a Young's modulus of 684 MPa for the 0.03 vol % graphene-loaded PVP nanocomposite (37 % enhancement); these experimental values agree closely with the theoretical predictions. The PVP-graphene interaction in the dispersion would indicate excellent load transfer after polymerization.

We utilized SEM to study the fracture surface in our graphene/PVP nanocomposites (Figure 9). Prior work has shown that graphene aggregates can be seen in SEM, but it is quite difficult to image individual dispersed graphene sheets in nanocomposites at low weight fractions via SEM [57]. We observed no signs of aggregation. Although we could not observe individual graphene sheets in the fracture surface of graphene/PVP nanocomposite, strong morphological changes were found due to the addition of only 0.03 vol % graphene in PVP. The graphene appears to inhibit crack formation in the nanocomposite as compared to the graphene-free sample. Even for only 0.03 vol % addition of graphene, the cracks on the surface of the nanocomposite are significantly reduced. Additional SEM images of the top surface of the composites are shown in Figure S5.

Electrical conductivity measurements of the PVP nanocomposites with and without graphene loading are shown in Table 3. The addition of 0.27 vol % of graphene in the PVP nanocomposite increases its electrical conductivity by 7 orders of magnitude, indicating that the percolation threshold lies below this value. These results compare favorably to other graphene-loaded electrically conductive nanocomposites [45].

The combination of SEM observations, conductivity measurements, and mechanical properties clearly indicates that the graphene stays dispersions during polymerization. Also, the fact that VP does not disperse graphene without the PVP stabilizer is interesting because VP and PVP have the same functional groups, which likely have a similar affinity for the graphene surface. However, the monomer is less effective in sterically hindering graphene reaggregation, whereas adsorbed PVP is kinetically trapped in its adsorbed state and prevents van der Waals contact between graphene sheets.

## 4. Conclusions

In conclusion, a simple method was shown for obtaining single-to-few layer pristine graphene directly from graphite by using PVP as a stabilizer. This technique allowed dispersion of pristine graphene sheets at high concentrations in a range of solvents with industrial relevance. The study indicated that the presence of PVP is critical for stabilizing the graphene dispersions. The dispersions were confirmed to be stable against lyophilization, pH changes, and temperatures greater than 100 $^{\circ}$C. Excellent enhancement in thermal conductivity of the fluid by addition of pristine graphene was also demonstrated. The obtained pristine graphene was utilized to make mechanically strong and electrically conductive graphene/PVP nanocomposites by bulk polymerization of VP. For a very low loading of graphene (0.03 vol %), the Young's modulus showed an increase of 37 % compared to PVP and the electrical conductivity of a 0.27 vol % loaded graphene/PVP nanocomposite was found to be seven orders of magnitude higher than PVP. These enhanced properties of the graphene/PVP nanocomposite indicate excellent load transfer between the graphene and the matrix and formation of a percolating network within the nanocomposite. The relatively high pristine graphene concentrations and excellent dispersion quality in wide range of solvents obtained from our method is promising for development of new graphene based materials. Potential applications include lightweight multifunctional nanocomposites and biomedical applications.

**Acknowledgements:**

We are thankful to Dr. Zhaoyang Fan for helpful comments on electrical conductivity measurements. We acknowledge Colin Young and Professor Matteo Pasquali of Rice University for their help with the Raman measurements. The SEM was performed at the TTU Imaging Center (funded by NSF MRI 04-511) supported by Dr. Mark J. Grimson and Professor Lauren S. Gollahon. We thank Abel Cortinas, Dr. Huipeng Chen, and Dr. Jun Zhao of TTU for helpful insights on dispersion processing and nanocomposite preparation. Funding was provided by the U.S. National Science Foundation (NSF) under award CBET-1032330 and by the Air Force Office of Scientific Research Young Investigator Program (AFOSR FA9550-11-1-0027).

**Supporting information available:**

More information about dispersion effectiveness, absorbance spectra, SEM, Raman, and mechanical testing are available in the supplementary information section.

| SOLVENTS | PVP-stabilized graphene concentration (mg/mL) | Extinction coefficient, $\alpha$ ($Lg^{-1}m^{-1}$) |
|---|---|---|
| WATER | 0.42 | 1293 |
| ETHANOL | 0.40 | 1940 |
| METHANOL | 0.40 | 1669 |
| DMF | 0.53 | 1602 |
| NMP | 0.63 | 1269 |
| DMSO | 0.45 | 2166 |
| VP | 0.72 | 1047 |

Table 1: By using PVP, single-to-few layer graphene was dispersed in various solvents without any chemical modification on graphene with relatively high concentrations. PVP concentration was 20 mg/mL in all cases. The extinction coefficients calculated based on the Lambert-Beer's law, are used to determine the concentrations of the further dispersions.

| Material | Thermal Conductivity (W/mK) |
|---|---|
| PVP/Ethanol | 0.18 |
| 0.018 vol % Graphene/PVP/Ethanol | 0.45 |

Table 2: Thermal conductivity of baseline fluid and graphene-loaded fluid. The graphene/PVP/ethanol dispersion was prepared by using an initial concentration of 10 mg/mL of PVP.

| Material | Electrical Conductivity (S/m) |
|---|---|
| PVP | $< 10^{-12}$ |
| 0.27 vol % Graphene/PVP | $2.6 \times 10^{-5}$ |

Table 3: Electrical conductivity results of baseline PVP and 0.27 vol % graphene/PVP composites, indicating percolation at 0.27 vol % graphene.

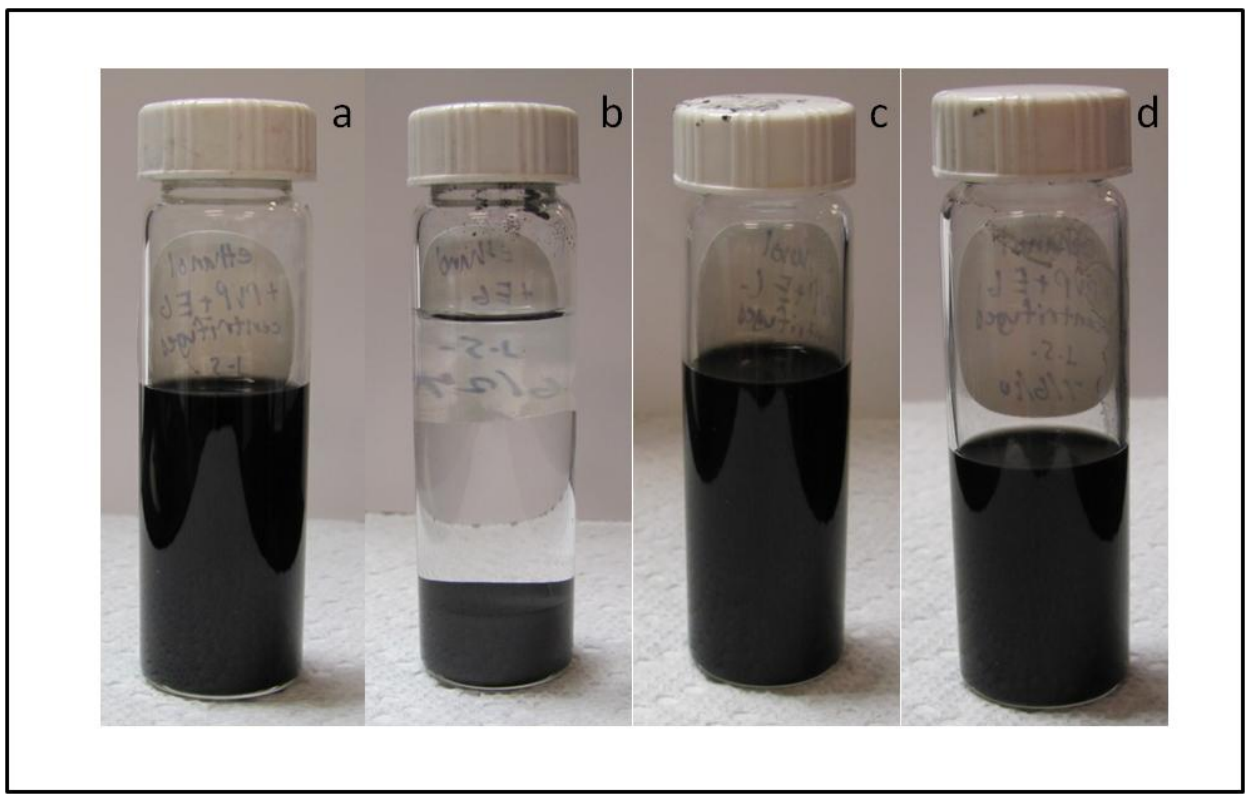

Figure 1: Digital photographs of pristine graphene in ethanol (a) after sonication and (b) after centrifugation and PVP-stabilized pristine graphene in ethanol (c) after sonication and (d) after centrifugation. Due to absence of any stabilizer, the graphene sheets aggregate due to van der Waals forces of attraction in (a) and (b). However, presence of PVP prevents aggregation and sedimentation of graphene in (c) and (d).

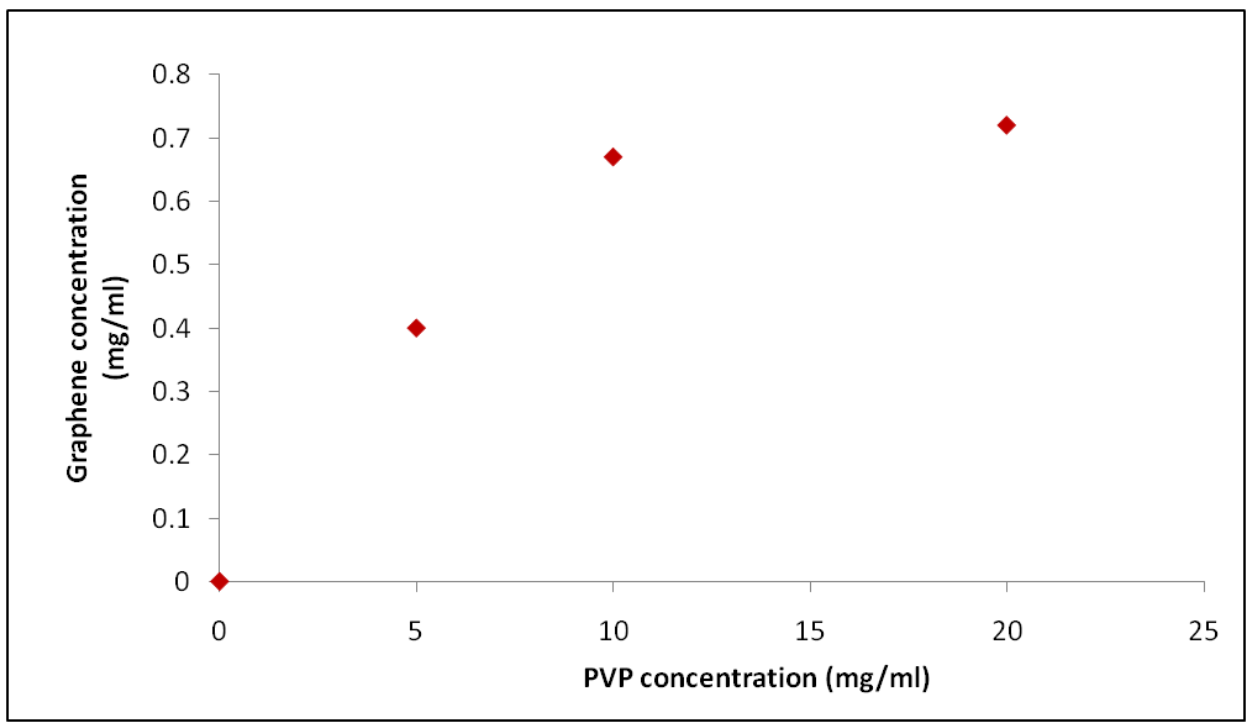

Figure 2: Study showing the effects of initial concentration of PVP on the final concentration of graphene dispersed in VP. With the increase in PVP concentration, the graphene concentration increases. However, beyond a PVP concentration of 10 mg/mL, the graphene concentration does not increase significantly.

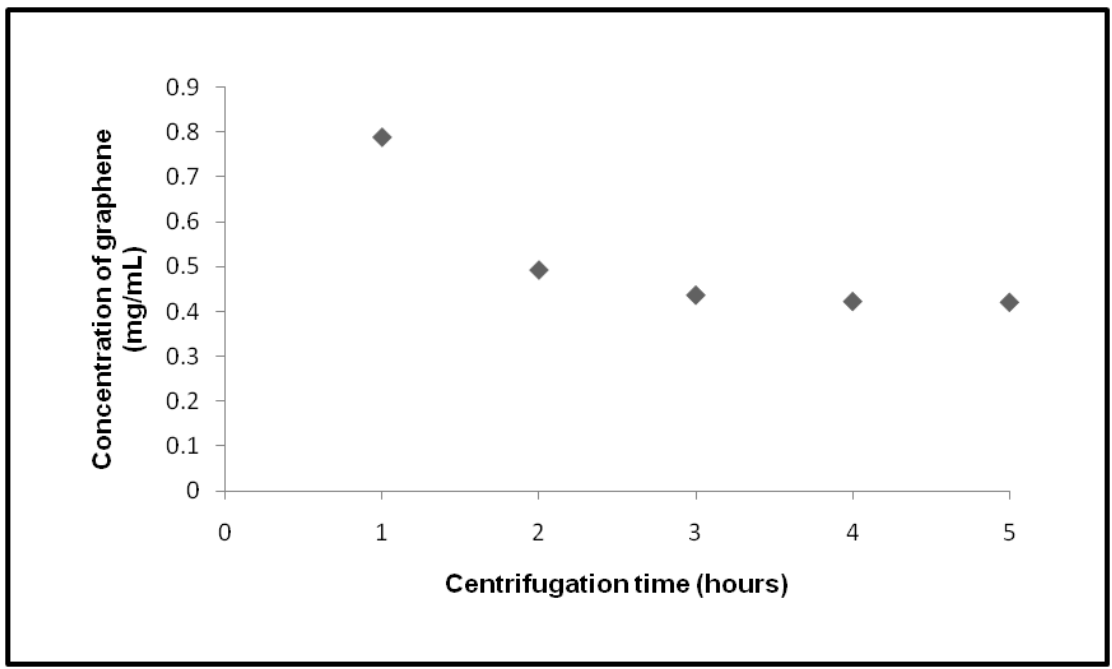

Figure 3: Study showing the effect of centrifugation time on the final concentration of PVP-stabilized graphene dispersed in water. The concentration of graphene stabilizes after 3 hours of centrifugation.

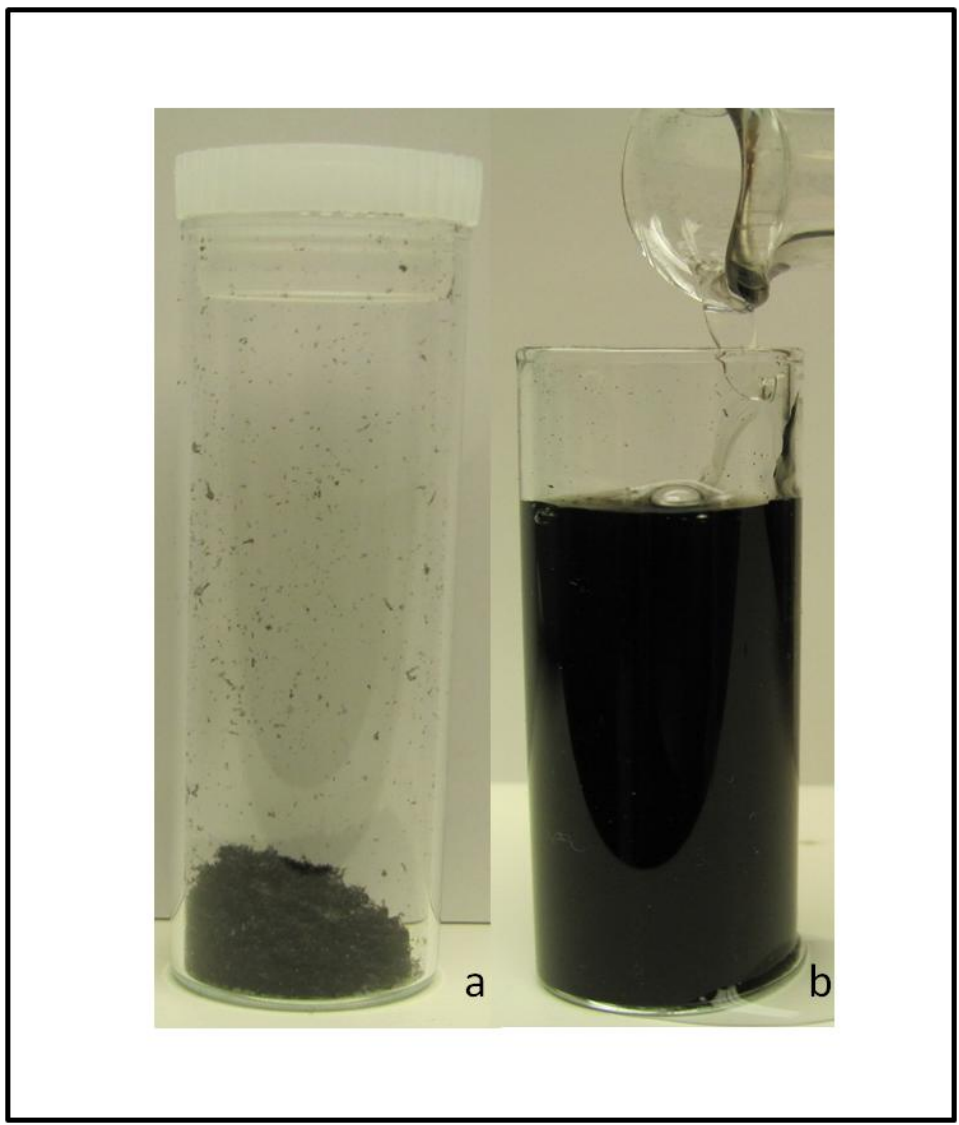

Figure 4: Digital photographs of (a) freeze-dried PVP stabilized graphene and (b) redispersed PVP stabilized graphene in water. The samples were readily dispersed in water after freeze-drying without the need of sonication and remain stable after further centrifugation.

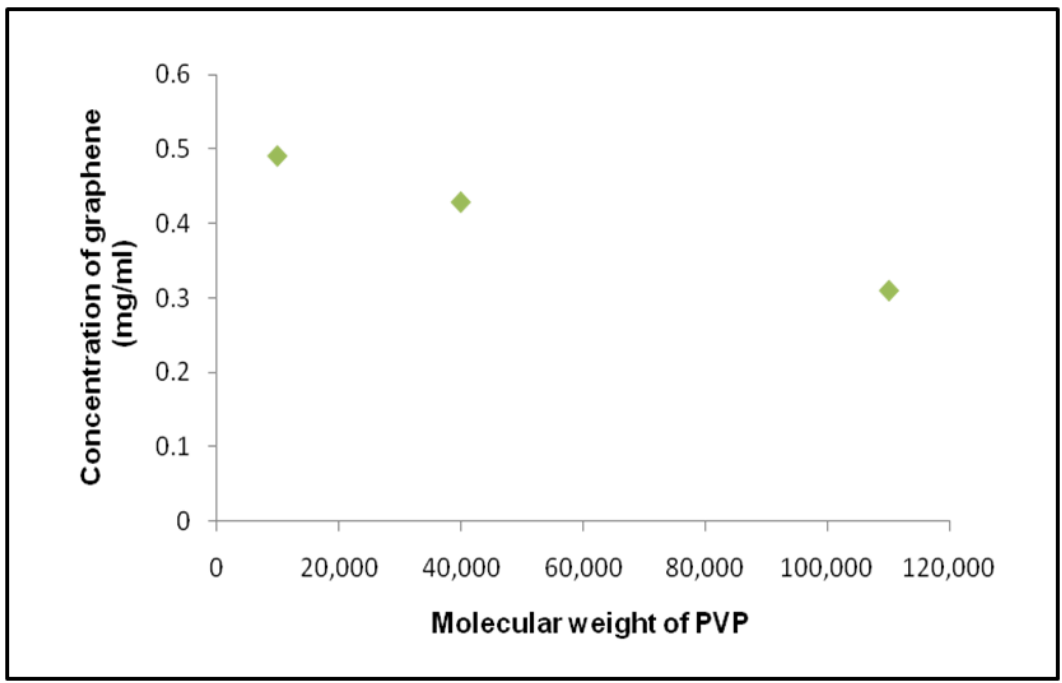

Figure 5: Study showing the impact of PVP molecular weight on the final concentration of graphene dispersed in VP. The initial concentration of PVP was kept constant at 10 mg/mL for all three molecular weights of PVP. Increase in the molecular weight of PVP leads to a decrease in the concentration of graphene in the dispersion.

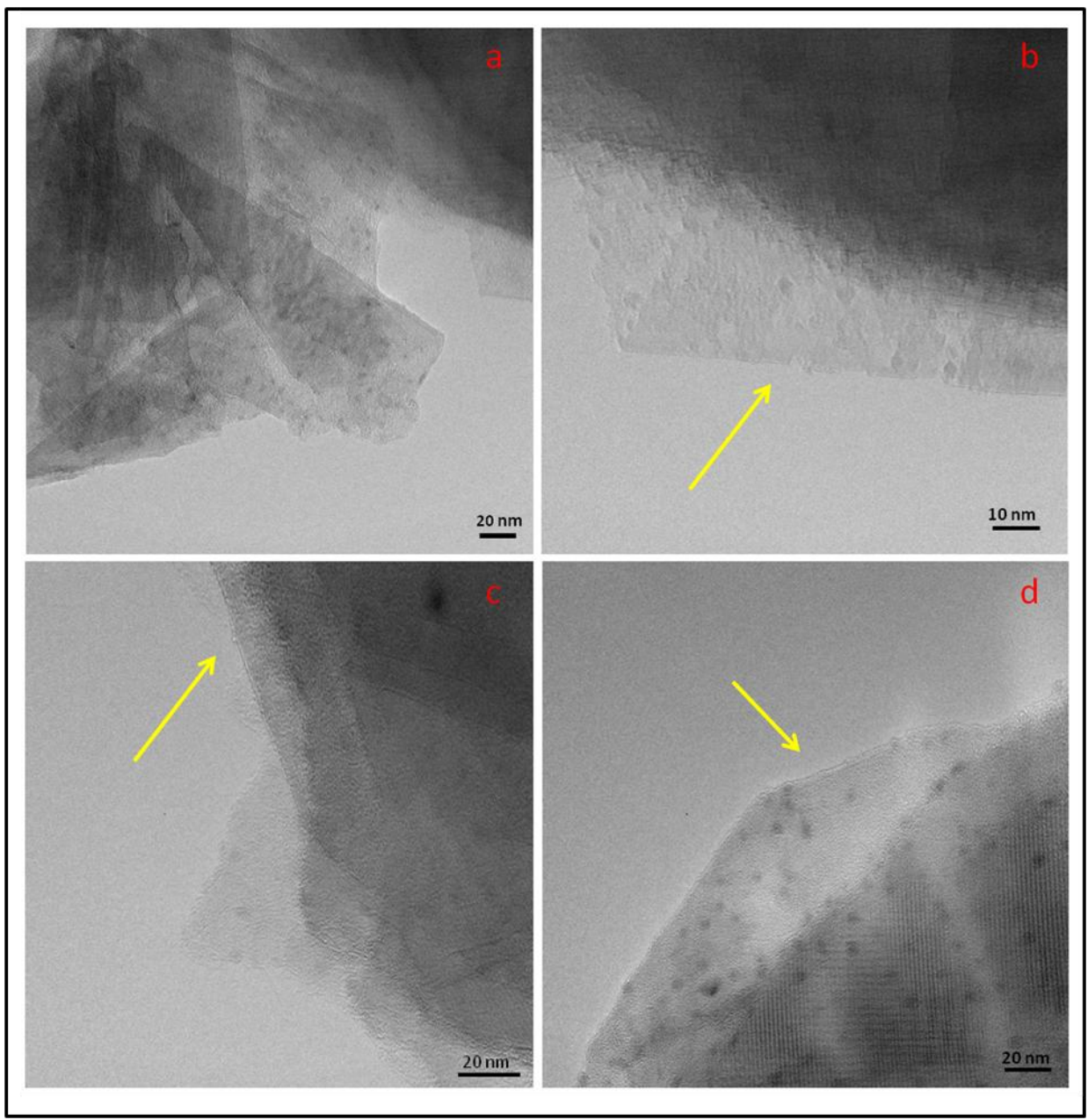

Figure 6: TEM images of PVP stabilized graphene deposited from ethanol. Fig 6 (a) shows folded graphene platelets. In Figures 6 (b), (c), and (d), on the edges of the graphene sheets, we observe that the polymer stabilized graphene is few layers thick (2-4 layers). Scale bars are 20 nm.

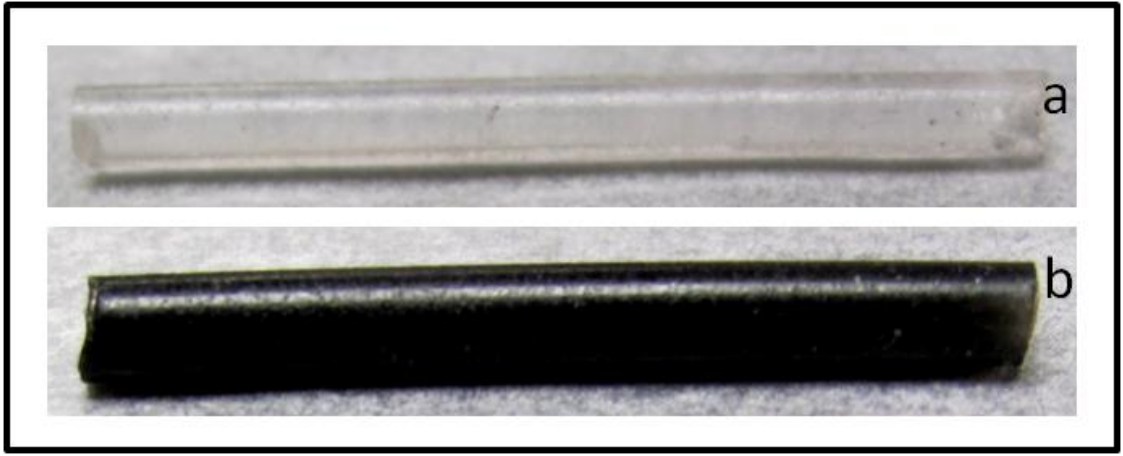

Figure 7: Digital photographs of composites made with UV light aided polymerization of (a) PVP (b) PVP stabilized graphene in PVP.

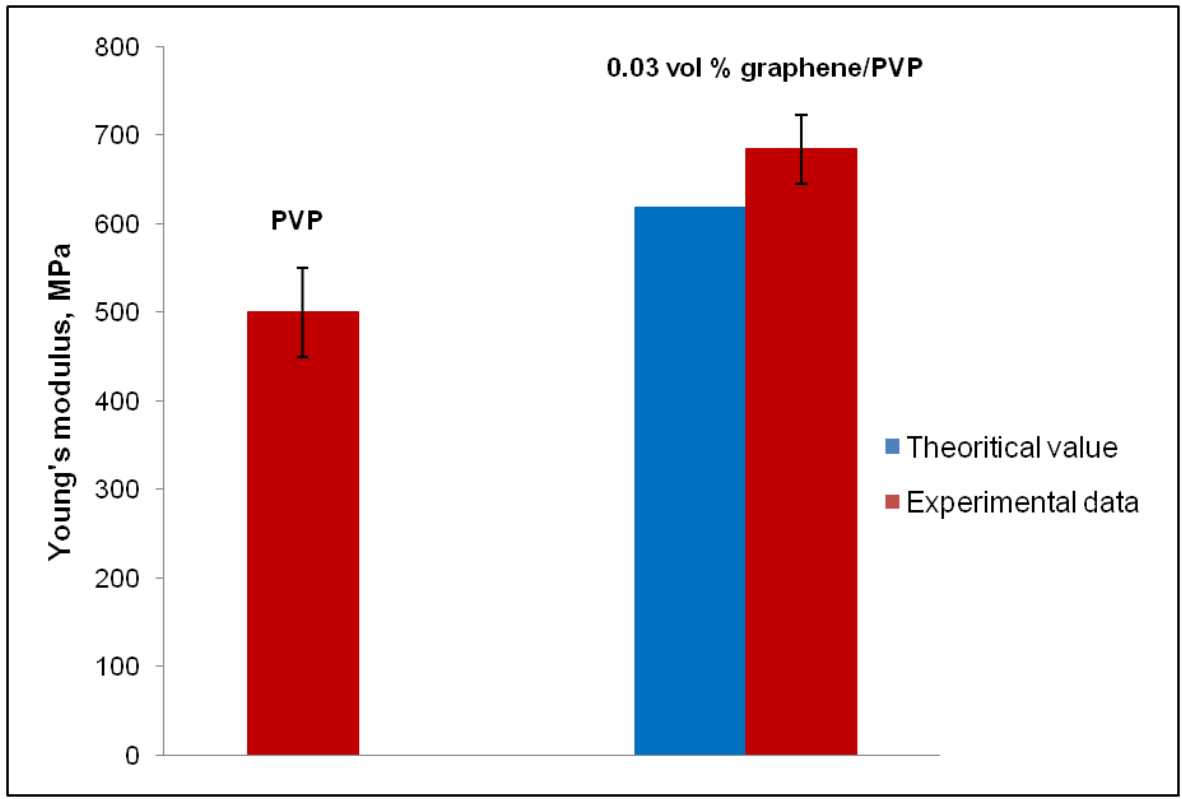

Figure 8: Young's Modulus of baseline PVP and PVP stabilized graphene/PVP composite sample with 0.03 vol % loading of graphene is compared. The samples were subjected to tensile loading, and the slope obtained from the stress-strain response determined the Young's modulus. The experimental value was also compared with the theoretical predictions obtained using the Halpin-Tsai theory for the nanocomposite sample.

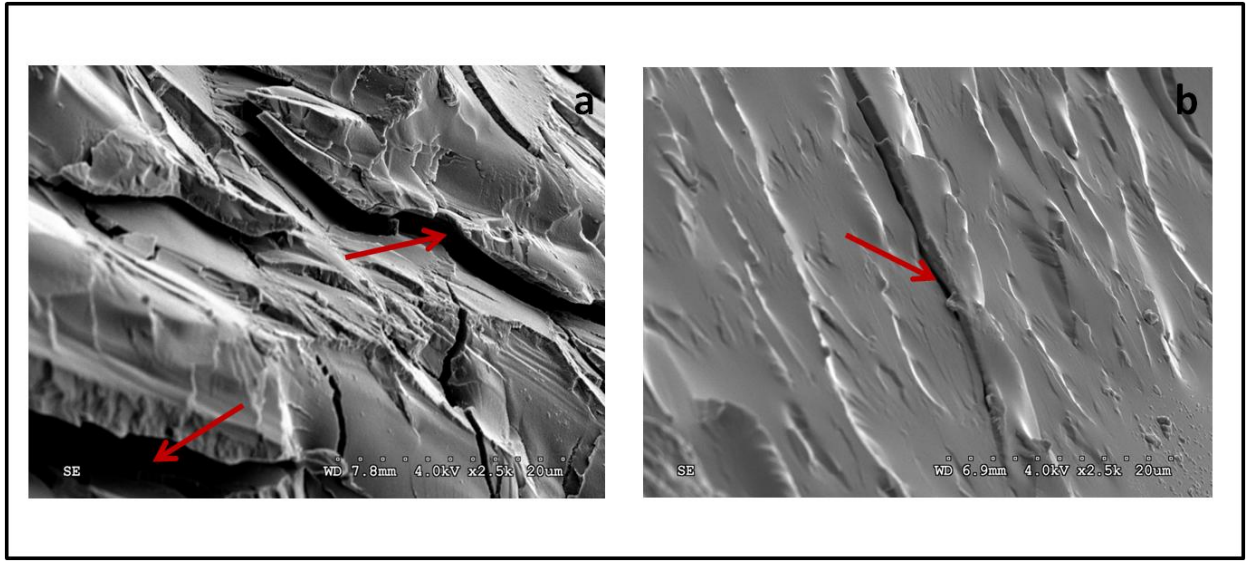

Figure 9: Scanning Electron Microscopy (SEM) images of (a) PVP fracture surface and (b) 0.03 vol % graphene/PVP fracture surface. The PVP fracture surface has large cracks indicated by the arrows in (a). The addition of graphene makes noticeable morphological change of the PVP surface; few cracks are observed on graphene/PVP fracture surface. Scale bars are 20 μm.

**Polymer-stabilized graphene dispersions at high concentrations in both aqueous and organic solvents for nanocomposite production**

Ahmed S. Wajid[1], Sriya Das[1], Fahmida Irin[1], H.S. Tanvir Ahmed[2], John L. Shelburne[1], Dorsa Parviz[1], Robert J. Fullerton[1], Alan F. Jankowski[2], Ronald C. Hedden[1], Micah J. Green[1*]

*[1]Department of Chemical Engineering, [2]Department of Mechanical Engineering, Texas Tech University, Lubbock, Texas 79409, USA*
*corresponding author: micah.green@ttu.edu*

# <u>Supplementary Information</u>

## <u>Section S1: Solvents tested</u>

The dispersibility of graphene was tested in 25 different solvents. All the solvents were used as received. The stabilizing polymer, PVP in our case, allows graphene to be dispersed in a variety of solvents that normally cannot disperse graphene.

.

| SOLVENTS | PVP SOLUBILITY | GRAPHENE DISPERSABILITY WITH PVP |
|---|---|---|
| Water | Soluble | Disperses |
| Ethanol | Soluble | Disperses |
| Acetone | Insoluble | Does not disperse |
| Chloroform | Soluble | Does not disperse |
| Dichloromethane | Insoluble | Does not disperse |
| Chlorobenzene | Insoluble | Does not disperse |
| Hexane | Insoluble | Does not disperse |
| Toluene | Insoluble | Does not disperse |
| Dimethyl Sulfoxide | Soluble | Disperses |
| Acetonitrile | Insoluble | Does not disperse |
| N,N Dimethylaniline | Insoluble | Does not disperse |
| N-Methylpyrrolidone | Soluble | Disperses |
| Tetrahydrofuran | Insoluble | Does not disperse |
| Methanol | Soluble | Disperses |
| Vinyl Pyrroldone | Soluble | Disperses |
| N,N Disopropyle ethylamine | Insoluble | Does not disperse |
| Ethyl acetate | Insoluble | Does not disperse |
| Dimethylformadide | Soluble | Disperses |
| Isopropanol | Soluble | Does not disperse |
| Pentane | Insoluble | Does not disperse |
| Vinyl acetate | Insoluble | Does not disperse |
| Hexanethyl | Insoluble | Does not disperse |
| Heptane | Insoluble | Does not disperse |
| Hexafluoroisopropanol | Soluble | Does not disperse |

Table S1:  The list of various solvents tested for solubility of PVP-stabilized graphene

**Section S2: Extinction coefficient measurement of graphene in different solvents**

The extinction coefficient of PVP stabilized graphene in various organic solvents was determined using the Lambert-Beer law, which states that the extinction coefficient of any substance varies linearly with the concentration. Thus by measuring absorbance of a dispersion at a wavelength of 660 nm and measuring the concentration of that dispersion (as determined by the vacuum filtration method described in the experimental section), we determined the concentrations of dispersions in that solvent by fitting a straight line fit the absorbance and the concentration of that sample in accordance with the theory. As expected for graphene, we do not see specific peaks in the visible region as we see in the case of single-walled carbon nanotubes.

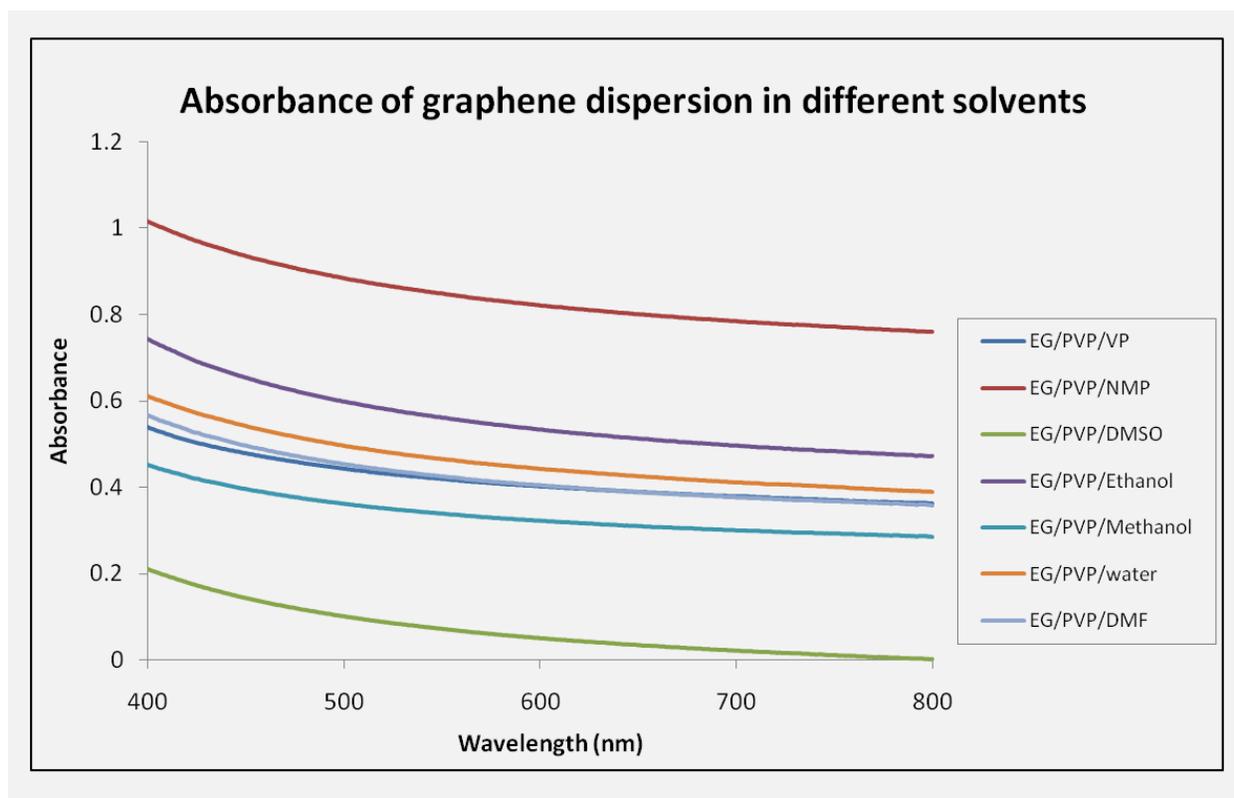

Figure S1: Absorbance spectra of PVP-stabilized graphene dispersed in various solvents such as Vinylpyrrolidone, N-Methylpyrrolidone, Dimethyl sulfoxide, Ethanol, Methanol, Dimethylformadide and water. (In all cases, the centrifuged sample is diluted by a factor of 4.)

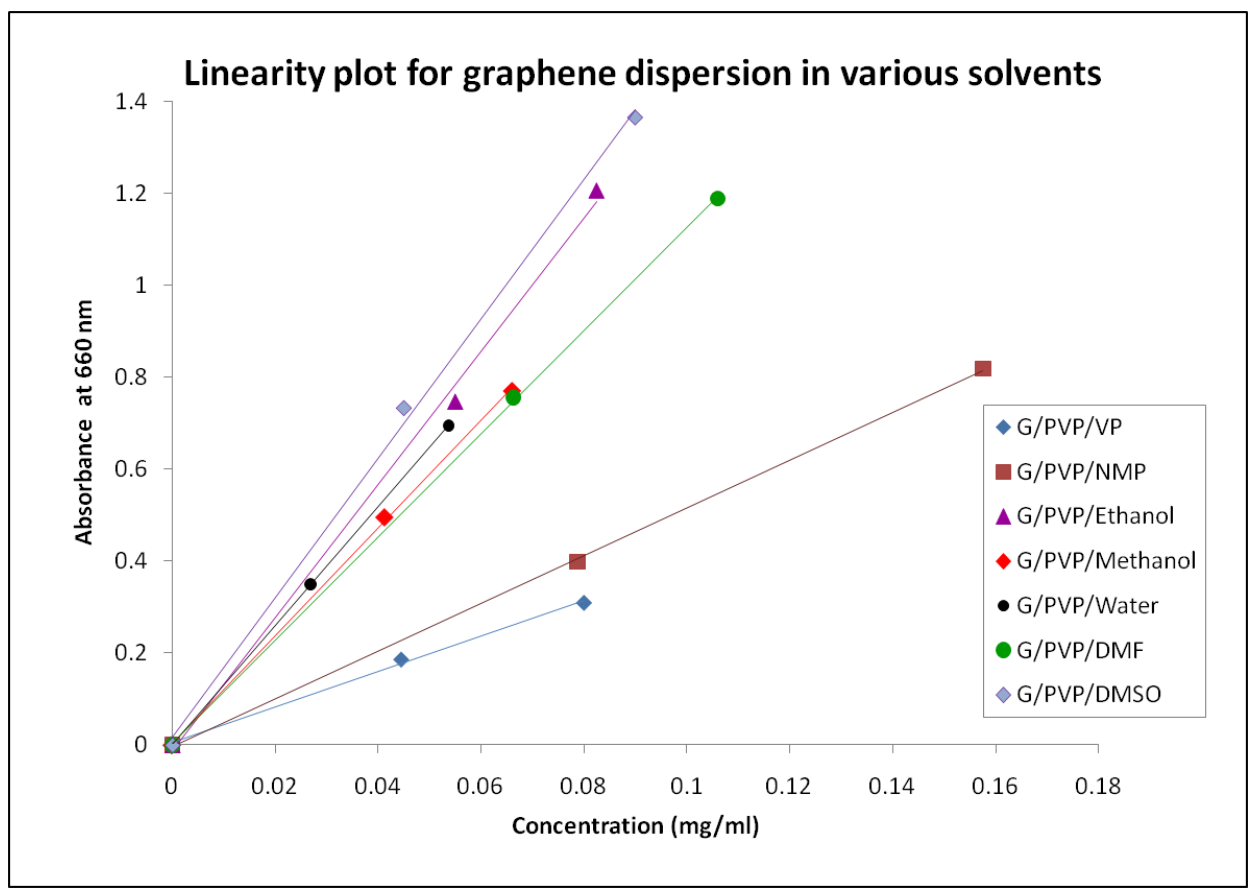

Figure S2: Linearity plot of PVP-stabilized graphene dispersed in various solvents.

**Section S3: Transmission Electron Microscopy (TEM) images**

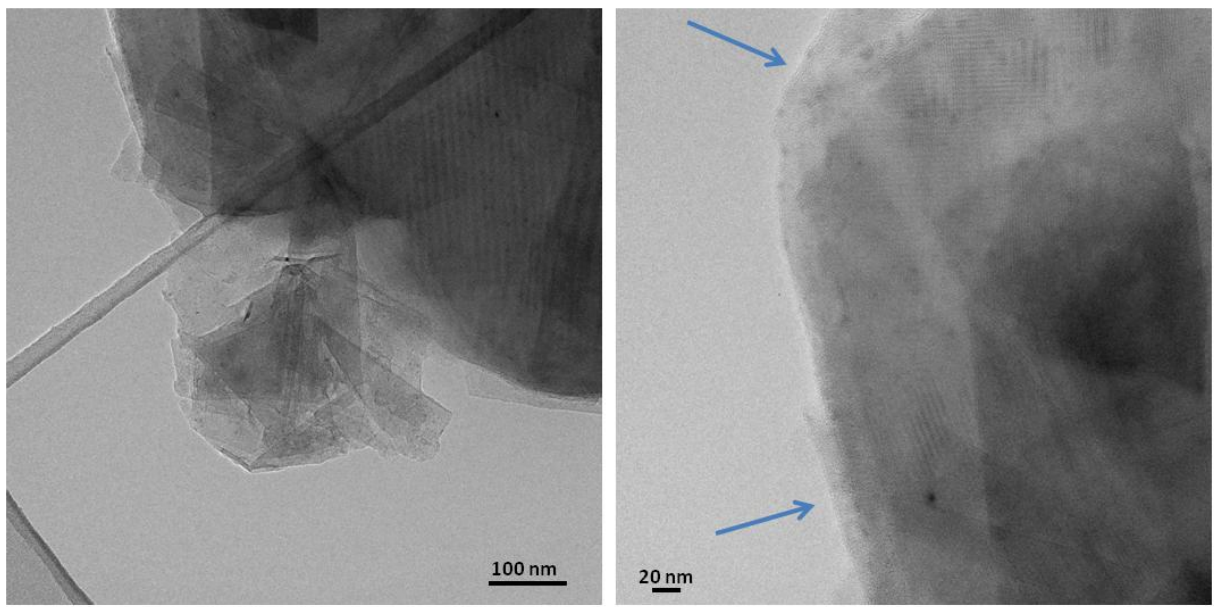

Figure S3: Additional TEM images of PVP stabilized graphene deposited from ethanol.

**Section S4: Raman spectroscopy**

Raman spectroscopy was used to characterize the degree of exfoliation of graphene in the solvent. In both of the above spectra we see a G-peak at ~ 1580 nm and a 2-D peak at ~ 2700 nm as expected for graphene. In the comparison of the Raman spectra of graphene and PVP-stabilized graphene dispersed in DMF we see that the intensity of the D- peak (~ 1460 $cm^{-1}$) increases as compared to the graphene source, EG. Prior work shows that the decrease in graphene flake size results in increase of graphene edges exposed per flakes [1]. These edges have $sp^3$ characteristics which cause the increase in intensity of the D- peak. We also see a decrease in the G-peak location of PVP-stabilized graphene by -3 $cm^{-1}$ [2].

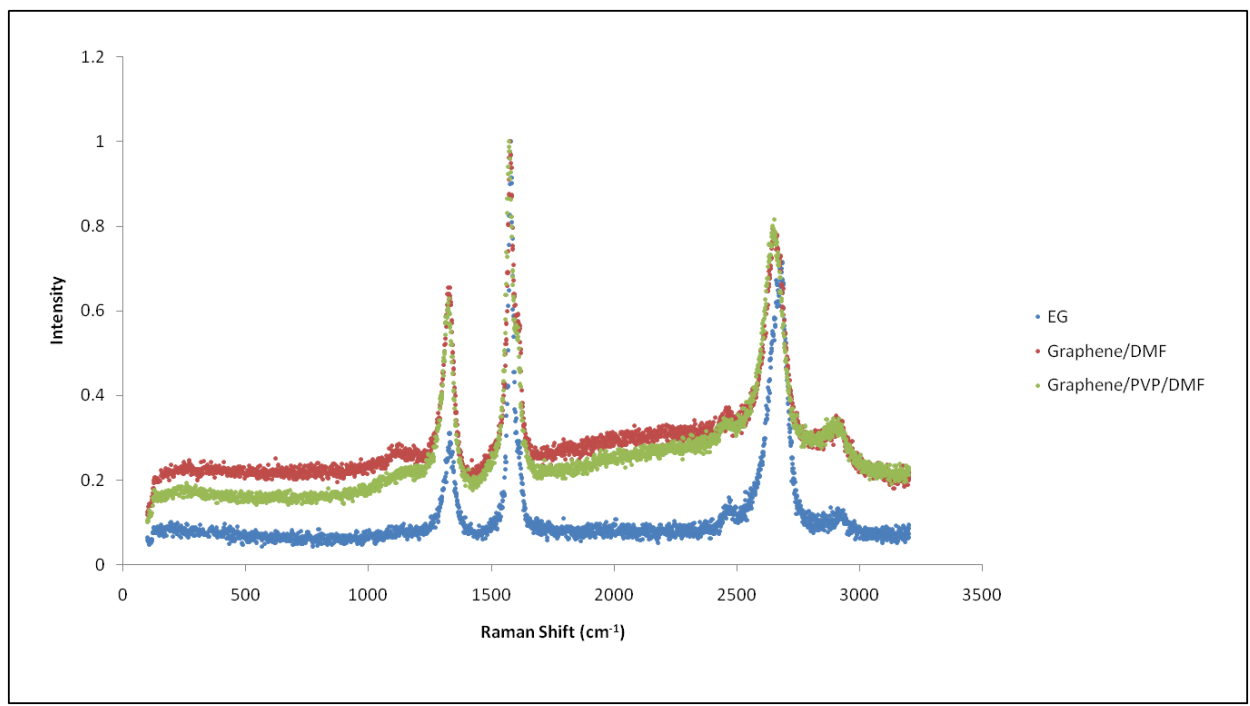

Figure S4: Comparison of Raman spectra of parent graphite (Blue), graphene dispersed in DMF (Green) and PVP-stabilized graphene dispersed in DMF (Red). The exfoliation of graphene in the case of graphene/PVP/DMF dispersion is confirmed by the shift in G-peak [2].

**Section S5: Scanning Electron Microscopy (SEM) images**

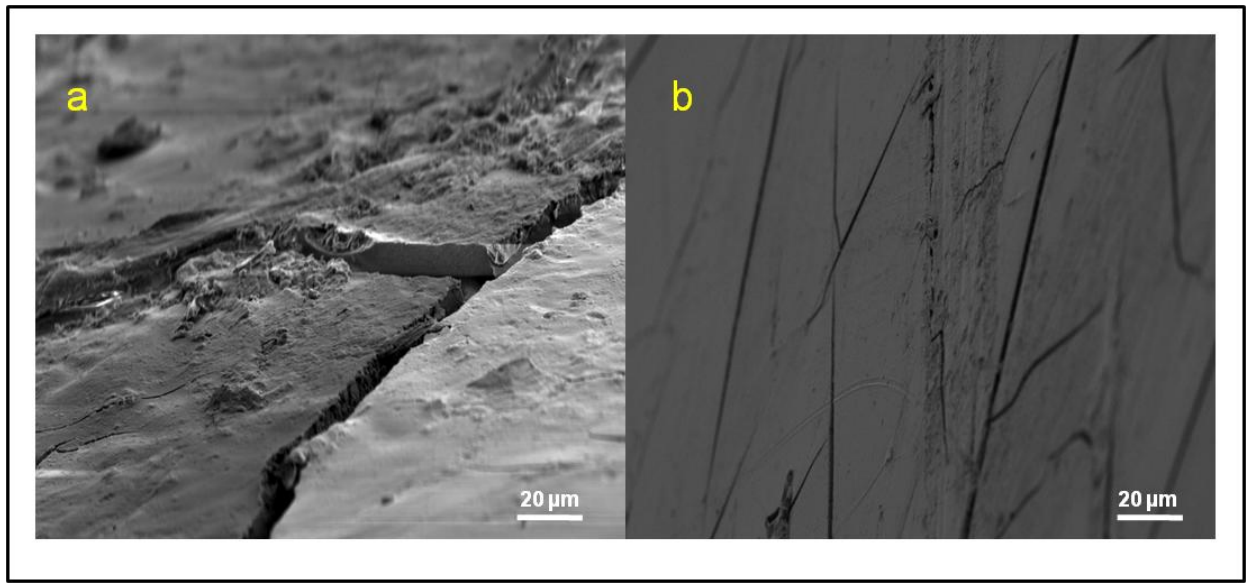

Figure S5: Scanning Electron Microscopy (SEM) images of (a) PVP top surface and (b) 0.03 vol % graphene/PVP top surface. The top surface of graphene/PVP nanocomposite shows relatively smaller cracks compared to the top surface of PVP.

**Section S6: Composite preparation**

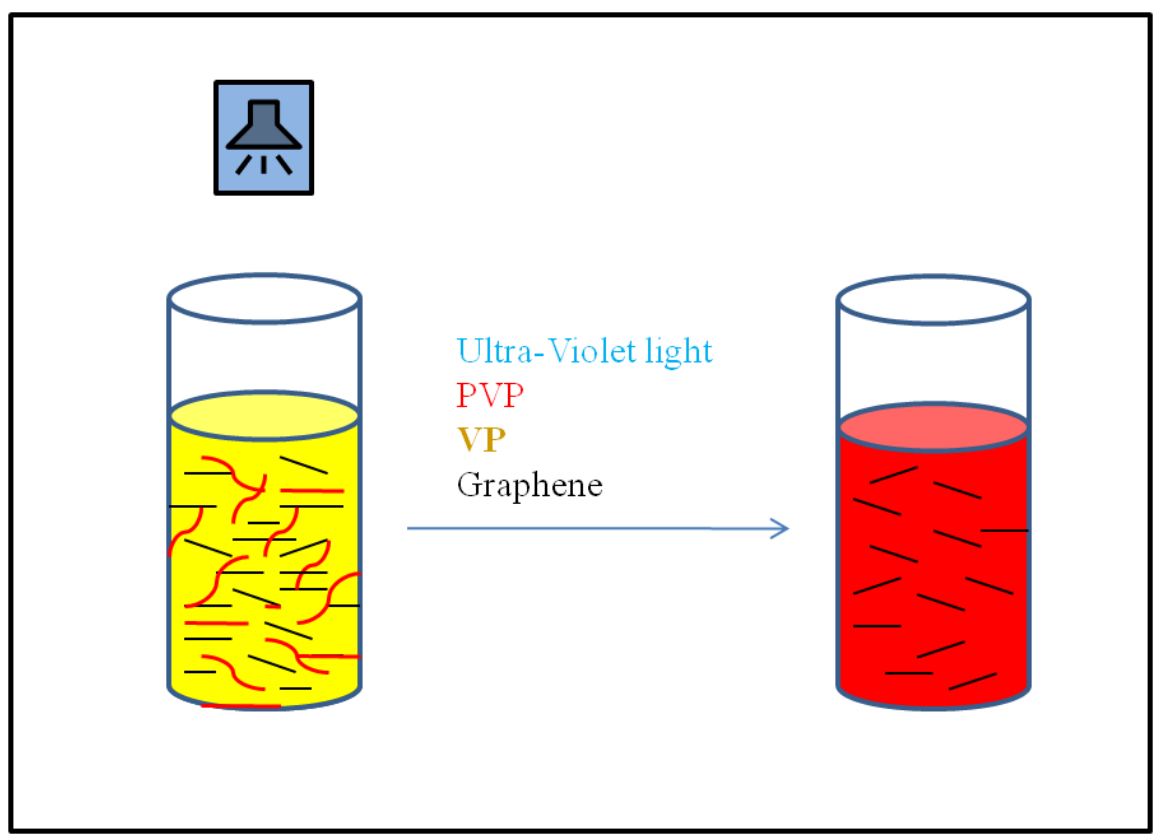

Figure S6: Schematic of UV-initiated polymerization of PVP-stabilized graphene in vinylpyrrolidone. This unique polymerization technique, where the stabilizing polymer is same as the matrix of the composite, results in excellent load transfer and mechanical property and mechanical property enhancement in the graphene-loaded PVP composite.

**Section S7: Theoretical Calculations of Tensile Properties**

We utilized the Halpin-Tsai model for randomly oriented filler-reinforced composites to predict the Young's Modulus of the nanocomposite [3-5]. The modified Halpin-Tsai equation used for our graphene/PVP composite is shown below.   (Note that in reference [3], Rafiee *et al.* incorrectly list two filler moduli when there is actually only one, $E_G$.)

$$E_c = \frac{3}{8} \frac{1 + \left(\frac{W+L}{t}\right)\left(\frac{\left(\frac{E_G}{E_{PVP}}\right) - 1}{\left(\frac{E_G}{E_{PVP}}\right) + \frac{W+L}{t}}\right) V_G}{1 - \left(\frac{\left(\frac{E_G}{E_{PVP}}\right) - 1}{(E_G / E_{PVP}) + (W+L)/t}\right) V_G} E_{PVP} + \frac{5}{8} \frac{1 + 2\left(\frac{(E_G/E_{PVP}) - 1}{(E_G/E_{PVP}) + 2}\right) V_G}{1 - \left(\frac{(E_G/E_{PVP}) - 1}{(E_G/E_{PVP}) + 2}\right) V_G} E_{PVP}$$

$E_c$ = Young's modulus of composite

$E_G$ = Young's modulus of graphene sheet ~ 1.01 TPa

$E_{PVP}$ =  Young's modulus of the PVP matrix ~ 500 MPa (experimental value)

$V_G$ = Volume content of graphene sheet ~ 0.0324 vol % = 0.000324

L, W, and t are the average graphene sheet length, width and thickness

L = 2.5 μm; W = 1.5 μm; t = 1.5 nm (These are the same values as used in [3], and these values are in the approximate range indicated by our TEM images.).

Substituting these values in the equation above, we get

$E_C$ = 619 MPa